\documentclass[a4paper,10pt,twocolumn,superscriptaddress,pra]{revtex4-1}
\usepackage[utf8]{inputenc}
\usepackage{mathtools}
\usepackage{amsmath}
\usepackage[left=1.5cm,top=2.2cm,right=1.5cm,bottom=2.2cm]{geometry}
\usepackage{relsize}
\usepackage{siunitx}
\usepackage{subfigure}
\usepackage{textcomp}
\usepackage{color}

\def\bra#1{\mathinner{\langle{#1}|}}
\def\ket#1{\mathinner{|{#1}\rangle}}
\def\be{\begin{equation}}
\def\ee{\end{equation}}
\def\ba{\begin{eqnarray}}
\def\ea{\end{eqnarray}}

\begin{document}

\title{Coherent exciton dynamics in a dissipative environment maintained by an off-resonant vibrational mode}
\author{E. K. Levi}
\affiliation{SUPA, School of Physics and Astronomy, University of St Andrews, KY16 9SS, UK}
\author{E. K. Irish}
\affiliation{School of Physics and Astronomy, University of Southampton, SO17 1BJ, UK}
\author{B. W. Lovett}
\affiliation{SUPA, School of Physics and Astronomy, University of St Andrews, KY16 9SS, UK}
\begin{abstract}
The interplay between an open quantum system and its environment can lead to both coherent and incoherent behaviour. We explore the extent to which strong coupling to a single bosonic mode can alter the coherence properties of a two-level system in a structured environment.
This mode is treated exactly, with the rest of the environment comprising a Markovian bath of bosonic modes. 
The strength of the coupling between the two-level system and the single mode is varied for a variety of different forms for the bath spectral density in order to assess whether the coherent dynamics of the two-level system are modified. We find a clear renormalisation of the site 
population oscillation frequency that causes an altered interaction with the bath. This leads to enhanced or reduced coherent behaviour of the two-level system depending on the form of the spectral density function. We present an intuitive interpretation, based on an analytical model, to explain the behaviour. 
\end{abstract}

\maketitle

\section{Introduction}\label{sec:intro}

Open quantum systems have been an active area of research for decades, but only recently have measurements been able to probe the dynamical effects of a coupled environment in detail\cite{Engel2007,Chin2012,Apollaro2011,Liu2011}. 
The environment has historically been thought to lead to deleterious decoherence of the open system, but it is now accepted that the role of the environment can be much more subtle than this.
In particular, the importance of structured environments in the dynamics and coherence of open quantum systems is now beginning to be recognised. 
For example, a series of recent theoretical \cite{Ritschel2011, Kreisbeck2012, Rey2013, Chin2013, Irish2014a, Killoran2014, Dijkstra2015,Nalbach2015} and experimental \cite{Lim2015, Novelli2015} studies have provided strong evidence that strongly coupled discrete molecular vibrations play a significant role in the speed, efficiency, and quantum coherence of energy transfer in photosynthetic and other molecular systems. 
In a different context, the high degree of control and precision possible in artificial nanosystems has enabled experimental measurement and engineering of noise spectral densities in condensed matter systems. 
Studies on micromechanical resonators have revealed a strongly sub-Ohmic spectral density (SD)\cite{Groblacher2015}; in another series of experiments, a superconducting flux qubit was used to probe the SD of a microwave transmission line, which could be tuned between Ohmic and Lorentzian forms using partial reflectors \cite{Haeberlein2015}.  

To mathematically model an open quantum system exactly one must fully capture the dynamics of the environment, and this is usually an impossible task: its Hilbert space is vast. Necessarily the environment must be treated using various approximations. One common description employs the Markov approximation, meaning that future behaviour depends only on the current state with no memory of preceding interaction\cite{Breuer2007}. Markovian environments are generally
straightforward to simulate since the dissipative system dynamics are characterised by a set of constant decay rates, which are found from the easily-obtained SD function.
When system-environment correlations decay more slowly the approximation becomes invalid and non-Markovian approaches are needed. There exists a spectrum of techniques for dealing with non-Markovianity to varying degrees including master equations\cite{Breuer2007}, quasi-adiabatic path 
integrals\cite{Makri1995,Makri1995a}, quantum Monte Carlo techniques\cite{Dalibard1992,Mak1994} and hierarchy equations of motion\cite{Tanimura1989}. 

In this paper we will introduce non-Markovianity straightforwardly by including part of the `environment' in the open system, that is our open system will consist of both the two-level system (TLS) whose dynamics we want to model, and a single oscillator mode (SM). The rest of the (bosonic) bath will be assumed to be weakly coupled to the open system and to be Markovian. Thus we are able to accurately study a structured environment consisting of a continuous background of modes with a strongly coupled single mode at one particular frequency. This framework maps to a host of physical 
situations: for example, a multi-site chromophore array in a protein bath typical of photosynthesis, where local protein vibrational modes interact strongly with excitons\cite{Kolli2012}, or
superconducting qubits coupled to nanomechanical oscillators or microwaves in resonators~\cite{OConnell2010,Wallraff2004}. We will present our work in the context of a two-site energy transfer system, or dimer, in the two-state subspace of single exciton levels. 


A single mode representation of certain strong features of a structured environmental coupling has been studied previously~\cite{Thorwart2004, Brito2008, Iles-Smith2014}, and symmetries have been exploited to reduce the complexity of the resulting system description\cite{Hossein-Nejad2010,Hossein-Nejad2014}.
However, the particular configuration of the TLS-SM-bath system investigated here is not represented in these earlier studies, in which the SM either couples to, or is not distinguished from, the bath. 
A similar model incorporating a single common mode was used to demonstrate enhanced coherence in the excitonic dynamics of an asymmetric dimer with individual Markovian baths for each dimer site\cite{Chen2014}. 
In contrast to our model, a polaron transform was used to approximately treat non-Markovian effects. 


In the next section we specify our TLS-SM-bath system, deriving a weak-coupling master equation to describe its dynamics. In Sec.~\ref{sec:prelim} we present numerical results for the dynamics of the TLS across a broad range of parameters. We analyse and discuss these findings in Sec.~\ref{sec:ana} and propose an intuitive and analytical interpretation of the complex dynamics observed. We conclude in Sec.~\ref{sec:conc}.

\section{Model}\label{sec:method}
 
\begin{figure}
 \centering
 \includegraphics[scale=0.45,keepaspectratio=true]{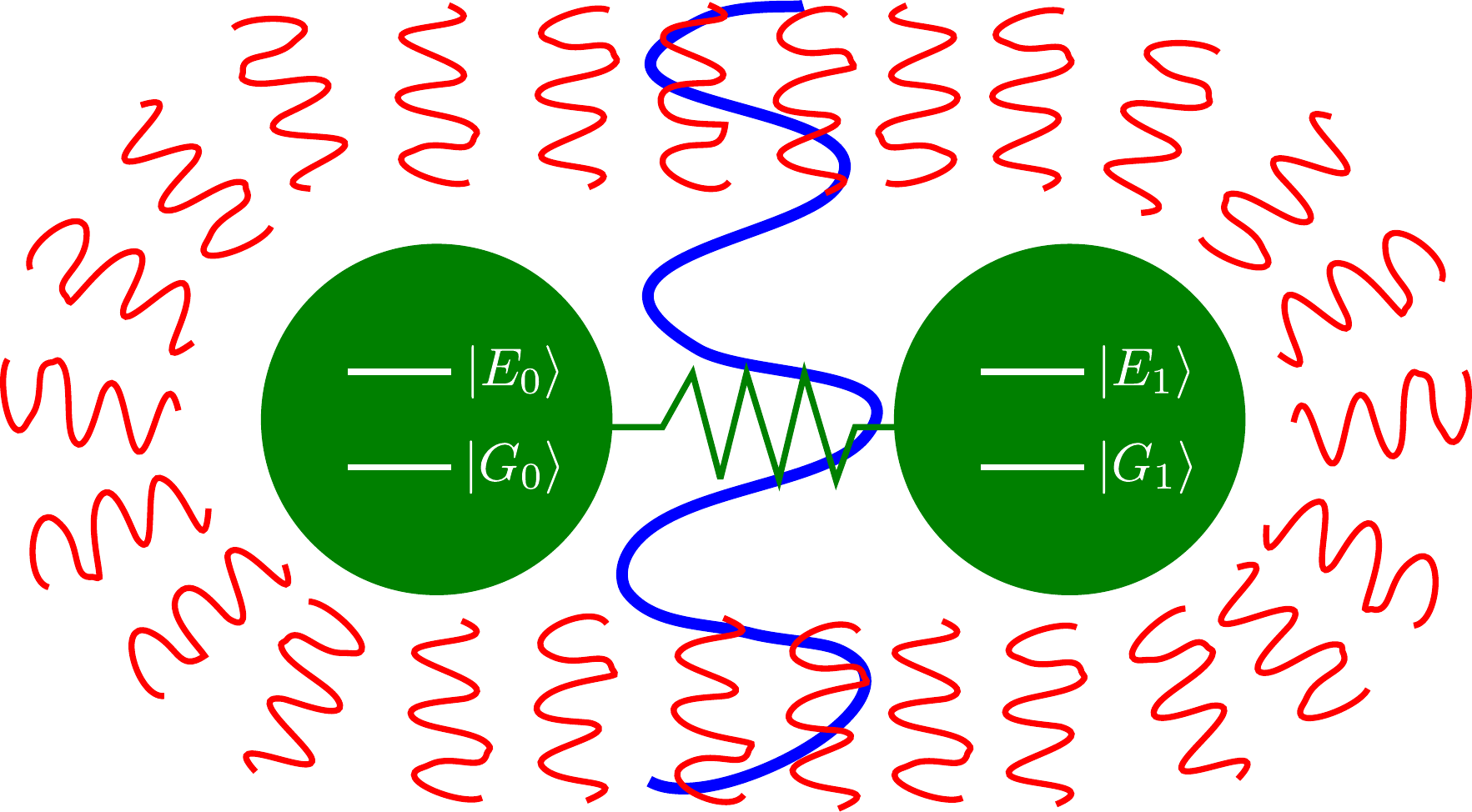}
 \caption{A schematic diagram of the model given in Eq.~\eqref{eq:underiv}. The symmetric dimer is composed of two identical sites with a F\"{o}rster-type interaction between them (green). This is linearly coupled to a Markovian bosonic environment of oscillators (red). One bosonic mode of the environment (blue) is assumed to be much more strongly coupled to the dimer than the others, and is therefore considered as part of the system that must be treated exactly.}
 \label{fig:schematic}
\end{figure}

The symmetric dimer consists of two identical sites, labelled `0' and `1', each with a ground $\ket{G}$ and single exciton excited state $\ket{E}$. Both sites are coupled to a single vibrational mode and a bosonic Markovian 
environment as displayed in Fig.~\ref{fig:schematic}. The total  Hamiltonian may be written
\be
\label{eq:underiv}
H_{\mathrm{Full}} = H_{\rm TLS}+H_{\rm SM}+H_{\rm TLS-SM}+H_{\rm B}+H_{\rm TLS-B}.
\ee
The electronic part of the dimer Hamiltonian is 
\be
H_{\rm TLS}=\varepsilon\left(\ket{E_0}\bra{E_0}+\ket{E_1}\bra{E_1}\right)- J(\ket{E_0 G_1}\bra{G_0 E_1}+{\rm H.c}),
\ee
where the excitation energy on each dimer site is $\varepsilon$ and the F\"{o}rster coupling between them is $J$. The terms involving the vibrational mode are:
\ba
H_{\rm SM}&= &\Omega\hat{a}^{\dagger}\hat{a};\\
H_{\rm TLS-SM} &=&-\left(g_0\ket{E_0}\bra{E_0}+g_1\ket{E_1}\bra{E_1}\right)\left(\hat{a}^\dagger+\hat{a}\right) ,
\ea
where the vibrational mode has frequency  $\Omega$, creation and annihilation operators $\hat{a}^{\dagger}$, $\hat{a}$ and 
couples to each site with strength $g_{0}=-g_{1}=g$. The terms involving the rest of the bosonic bath are:
\ba
H_{\rm B}	&=& \sum_{\mathbf{q}} \omega_{\mathbf{q}}\hat{b}_{\mathbf{q}}^{\dagger}\hat{b}_{\mathbf{q}}; \\
H_{\rm TLS-B}&=&-\sum_{\mathbf{q}}\left(h_{\mathbf{q},0}\ket{E_0}\bra{E_0}+h_{\mathbf{q},1}\ket{E_1}\bra{E_1}\right)\left(\hat{b}_{\mathbf{q}}^{\dagger}+\hat{b}_{\mathbf{q}}\right).\nonumber\\ 
\ea
This environment is comprised of harmonic modes of wave vector $\mathbf{q}$ with frequencies $\omega_{\mathbf{q}}$, creation and annihilation operators 
$\hat{b}^{\dagger}_{\mathbf{q}}$, $\hat{b}_{\mathbf{q}}$ and coupling strengths of $h_{\mathbf{q},0}=-h_{\mathbf{q},1}=h_{\mathbf{q}}$. 

The Hamiltonian in Eq.~\ref{eq:underiv} consists of three uncoupled subspaces with basis vectors
$\{|G_{0}G_{1}\rangle\}$, $\{|E_{0}E_{1}\rangle\}$ and $\{|E_{0}G_{1}\rangle,\,|G_{0}E_{1}\rangle\}$. In order to study energy transfer dynamics we may focus only on the last: the two-dimensional single excitation subspace. To simplify notation we then define $|0\rangle=|E_{0},G_{1}\rangle$ and $|1\rangle=|G_{0},E_{1}\rangle$.

In this subspace our Hamiltonian becomes:
\begin{equation}\label{eq:SBM}
H_{\rm F}=-J\hat{X}-\hat{Z}g(\hat{a}^{\dagger}+\hat{a})-\hat{Z}\displaystyle\sum_{\mathbf{q}}h_{\mathbf{q}}(\hat{b}_{\mathbf{q}}^{\dagger}+\hat{b}_{\mathbf{q}})+\Omega\hat{n}+\displaystyle\sum_{\mathbf{q}}\omega_{\mathbf{q}}\hat{n}_{\mathbf{q}}, 
\end{equation}
where $\hat{n}=\hat{a}^{\dagger}\hat{a}$ and $\hat{n}_{\mathbf{q}}=\hat{b}_{\mathbf{q}}^{\dagger}\hat{b}_{\mathbf{q}}$ are the number operators for the SM and the ${\mathbf{q}}$-th bath mode. 
We have now reformulated the problem into that of a single TLS interacting with the SM and environment where the site basis Pauli matrices are $\hat{Z}=|0\rangle\langle0|-|1\rangle\langle1|$ and $\hat{X}=|0\rangle\langle1|+|1\rangle\langle0|$. 
The form in Eq.~\ref{eq:SBM} is that of a modified spin-boson Hamiltonian; the spin-boson model is one of the most utilised and investigated descriptions of open quantum system behaviour\cite{Weiss2008,Leggett1987,Wilhelm2004,Ishizaki2009a}. 
Whilst not exactly solvable (except in special cases) it contains information about the interplay between a two-level `spin' system and a harmonic bath. 
Our model differs from the standard spin-boson model by the addition of the SM to the part of the system that is treated exactly. 
Both the SM and the bath couple to the TLS via a linear displacement with respect to the site basis.
 
Disregarding the bath for now, Eq.~\ref{eq:SBM} becomes:
\begin{equation}\label{eq:singleSBM}
H=-J\hat{X}-\hat{Z}g(\hat{a}^{\dagger}+\hat{a})+\Omega\hat{n}. 
\end{equation}
This is amenable to the Fulton-Gouterman transformation (FGT) which can simplify its solution. First introduced in 1961\cite{Fulton1961} and later refined and extended\cite{Wagner1984a}, the FGT exploits the parity 
symmetries of the system to diagonalise the Hamiltonian in the TLS subspace. It is a unitary transformation and, although there are various equivalent forms in the literature we shall use\cite{Paganelli2006} 
\begin{equation}
U=\frac{1}{\sqrt{2}}\biggl[|0\rangle\langle0|-|1\rangle\langle0|+\hat{P}\bigl(|0\rangle\langle1|+|1\rangle\langle1|\bigr)\biggr].
\end{equation}
The oscillator parity operator, $\hat{P}=(-1)^{\hat{n}}$, obeys the anti-commutation relations
\begin{equation}
\{\hat{P},\hat{a}\}=\{\hat{P},\hat{a}^{\dagger}\}=0.
\end{equation}
 
The function of the FGT is to isolate states of the same parity, which do not couple to states of opposite parity. As shown in Appendix~\ref{app:FGT}, the FGT 
is equivalent to a change of basis and state reordering. Application of the FGT to Eq.~\ref{eq:singleSBM} yields
\begin{equation}
\tilde{H}=UHU^{\dagger}=\frac{H}{2}^{+}|0\rangle\langle0|+\frac{H}{2}^{-}|1\rangle\langle1|,
\end{equation}
thus creating subspace Hamiltonians of conserved parity:
\begin{equation}\label{eq:FGTsubs}
H^{\pm}=\Omega\hat{n}-g(\hat{a}^{\dagger}+\hat{a})\mp J\hat{P}.
\end{equation}
The eigenstates of these transformed Hamiltonians, expanded in the site basis, are
\begin{equation}\label{eq:FGstates}
|\psi_{k}^{\pm}\rangle=\frac{1}{\sqrt{2}}\bigl[\pm|0\rangle+\hat{P}|1\rangle\bigr]|\phi^{\pm}_{k}\rangle,
\end{equation}
where $|\phi^{\pm}_{k}\rangle$ are the eigenstates of the excitation parity subspace Hamiltonians. To clarify, the Schr\"{o}dinger equations are then $H^{\pm}|\phi^{\pm}_{k}\rangle=E^{\pm}_{k}|\phi^{\pm}_{k}\rangle$ and 
$\tilde{H}|\psi_{k}^{\pm}\rangle=E^{\pm}_{k}|\psi_{k}^{\pm}\rangle$.
 
Using a Fock basis for the SM and $\ket{\pm}\equiv(\ket{0}\pm\ket{1})/\sqrt{2}$ for the TLS, the FGT shows that the Hamiltonian takes a tridiagonal form that can be diagonalised efficiently. It is of course necessary to truncate the expansion of the SM, but for all calculations we ensure that simulations are numerically converged.
With this in mind we shall refer to our 
treatment of this TLS-SM system as exact. 
 
We now re-introduce the bath. In the  $|\pm\rangle$ basis, the Hamiltonian in Eq.~\ref{eq:singleSBM} becomes
\begin{equation}
H_{\rm S}=-J\hat{\sigma}_{z}-\hat{\sigma}_{x}g(\hat{a}^{\dagger}+\hat{a})+\Omega\hat{n},
\end{equation}
where now $\hat{\sigma}_{z}=|+\rangle\langle+|-|-\rangle\langle-|$, $\hat{\sigma}_{x}=|+\rangle\langle-|+|-\rangle\langle+|$, and we have introduced the subscript `S' to denote the system of TLS and SM. The Hamiltonian describing the interacting system and bath is then 
\begin{equation}\label{eq:hamil}
H_{\rm F}=H_{\rm S}+\displaystyle\sum_{\mathbf{q}}\omega_{\mathbf{q}}\hat{n}_{\mathbf{q}}+\hat{\sigma}_{x}\displaystyle\sum_{\mathbf{q}}h_{\mathbf{q}}\bigl(\hat{b}_{\mathbf{q}}^{\dagger}+\hat{b}_{\mathbf{q}}\bigr).
\end{equation} 
From this Hamiltonian we can derive a Markovian master equation, assuming that the TLS-SM system is coupled only weakly to the environmental oscillators. We have found a convenient form for the eigenstates of 
$H_{\rm S}$ which simplifies the master equation derivation if we describe the $\sigma_{x}$ interaction in this basis; this is shown in Appendix~\ref{app:interact}. 
 
The general form of the master equation following the Born-Markov approximation is, in the interaction picture:
\begin{equation}\label{eq:MEgen}
\frac{\mathrm{d}\rho_{\rm S}(t)}{\mathrm{d}t}=-\int\limits_{0}^{\infty}\mathrm{d}s \: \mathrm{Tr}_{\rm B}\Bigl[H_{I}(t),\bigl[H_{I}(t-s),\rho_{\rm S}(t)\otimes\rho_{\rm B}\bigr]\Bigr],
\end{equation}
which describes the dynamics of a  reduced system density matrix, $\rho_{\rm S}(t)$. $\rho_{\rm B}$ is the bath density matrix, assumed to represent a thermal state at all times, and $H_{I}(t)$ is the interaction-picture interaction Hamiltonian\cite{Breuer2007}. For our 
problem this becomes:
\begin{widetext}
\begin{align}\label{eq:withSA}
\frac{\mathrm{d}\rho _{\rm S}(t)}{\mathrm{d}t}=\displaystyle\sum_{k,k'}\biggl[ & \Bigl(\Gamma(-\Lambda_{kk'})+\Gamma'(\Lambda_{kk'})\Bigr)\Bigl(2\hat{\zeta}^{-}_{k'k}\rho _{\rm S}(t)\hat{\zeta}^{+}_{kk'}-\hat{\zeta}^{+}_{kk'}\hat{\zeta}^{-}_{k'k}\rho _{\rm S}(t)-\hat{\zeta}^{-}_{k'k}\hat{\zeta}^{+}_{kk'}\rho _{\rm S}(t)\Bigr) \nonumber \\
+ & \Bigl(\Gamma(\Lambda_{kk'})+\Gamma'(-\Lambda_{kk'})\Bigr)\Bigl(2\hat{\zeta}^{+}_{kk'}\rho _{\rm S}(t)\hat{\zeta}^{-}_{k'k}-\hat{\zeta}^{-}_{k'k}\hat{\zeta}^{+}_{kk'}\rho _{\rm S}(t)-\hat{\zeta}^{+}_{kk'}\hat{\zeta}^{-}_{k'k}\rho _{\rm S}(t)\Bigr)\biggr].
\end{align}
\end{widetext}
The $\hat{\zeta}^{\pm}_{ij}$ operators are Lindblad operators describing transitions between the FGT subspaces, from eigenstate $j$ to eigenstate $i$; these are defined explicitly in 
Eqs.~\ref{eq:fgtswitch1}~and~\ref{eq:fgtswitch2}. The transition rates are governed by $\Gamma$ and $\Gamma'$ which are defined in Eqs.~\ref{eq:startgam}-\ref{eq:endgam}; these rely on bath operator 
correlators and by extension the SD $\chi(\omega)=\sum_{\mathbf{q}}|h_{\mathbf{q}}|^{2}\,\delta(\omega-\omega_{\mathbf{q}})$ and they are derived in Appendix~\ref{app:maseq}. $\Lambda_{kk'}$ is a difference between 
FGT subspace eigenvalues, $E^{+}_{k}-E^{-}_{k'}$. We used numerical simulations\cite{Johansson2013} to solve Eq.~\ref{eq:withSA}, the full derivation of which can be found in Appendix~\ref{app:maseq}. 

We will use the following SD:
\begin{equation}\label{eq:ohmlike}
\chi_{m}(\omega)=A_m\omega^m\mathrm{e}^{-\omega^2/\omega_{m,c}^2},
\end{equation}
where varying $m$ moves between an Ohmic ($m=1$) and super-Ohmic ($m>1$) form. The Gaussian cut-off in this definition ensures that $\chi\rightarrow0$ for $\omega\gg\omega_{m,c}$, the cut-off frequency, meaning high 
frequency modes do not contribute to the dynamics. We impose $\chi(\omega)=0$ for $\omega\leq0$ and we introduce $\omega_{p}$ to denote the frequency at the peak of the SD, which is related to $m$ and $\omega_{m, c}$. For comparison we will also sometimes use 
a Lorentzian-like form for the SD:
\begin{equation}\label{eq:lorentz}
\chi_{L}(\omega)=\frac{A_L\omega W^{2}}{(\omega-\omega_{L,c})^{2}+W^{2}},
\end{equation}
where $W$ is the half width at half maximum and $\omega_{L,c}$ is determined by fixing $\omega_{p}$. The normalisation factors, $A_m$ and $A_L$, in Eqs.~\ref{eq:ohmlike}~and \ref{eq:lorentz} are related to a property of the bath known as the reorganisation energy:
\begin{equation}\label{eq:reorg}
\lambda=\int\limits_{0}^{\infty}\mathrm{d}\omega\frac{\chi(\omega)}{\omega}.
\end{equation}
This quantifies the energy associated with the bath as it interacts with the TLS\cite{McCutcheon2011a}. Defining a fixed value of $\lambda$ ensures fair SD comparisons in simulations; normalisation factors can then be calculated 
using Eq.~\ref{eq:reorg}. We also fix $\omega_{p}$ to ensure peak alignment during comparisons. 

\begin{figure}
 \centering
 \includegraphics[scale=0.45,keepaspectratio=true]{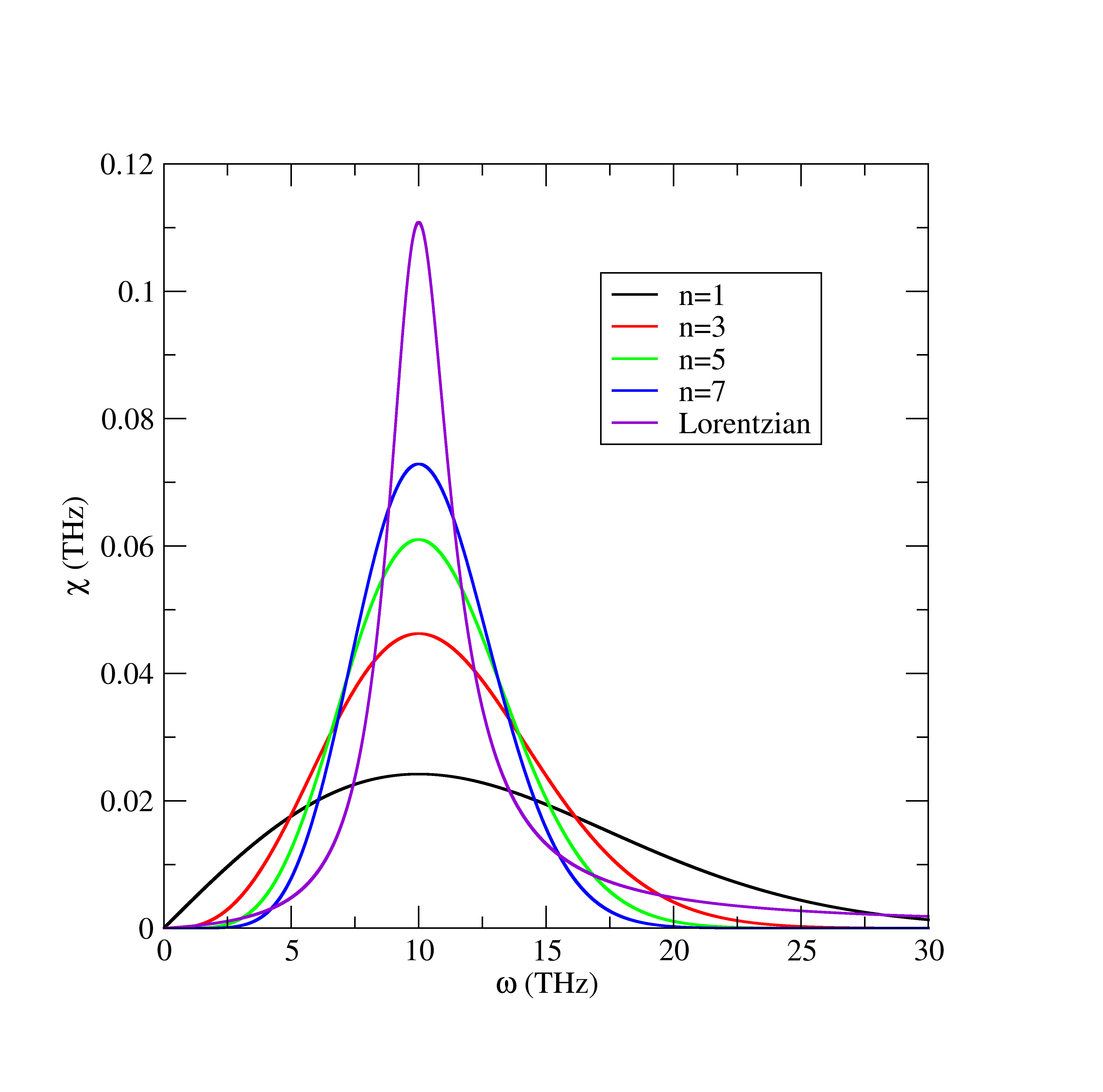}
 \caption{A comparative plot of the five spectral densities considered here, with $\lambda=0.05$~\si{\tera\hertz} and $\omega_{p}=10$~\si{\tera\hertz}. The Ohmic SD is given by $m=1$, while $m=3,5,7$ represent increasingly peaked super-Ohmic spectral densities. The Lorentzian SD has width $W=1.5$~\si{\tera\hertz}.}
 \label{fig:specdens}
\end{figure}

 
We have studied dynamics for $m=1,\,3,\,5,\,7$ and Lorentzian spectral densities; for reference these are shown (for a constant $\lambda$ and $\omega_{p}$) in Fig.~\ref{fig:specdens}. Ohmic spectral densities are frequently used for 
 low temperatures or for surface-surface tunnelling problems, while super-Ohmic descriptions apply well to bulk phonon baths\cite{Weiss2008,Louis1995}. The Lorentzian SD in Fig.~\ref{fig:specdens} 
is far more peaked than the Ohmic and super-Ohmic curves, and is a simple and physical analytic tool for investigating structured spectral densities with sharp features. 

Before we move on to discussing results in Sec.~\ref{sec:prelim} we will introduce our initial conditions. There are two straightforward ways of initialising a thermal SM: 1) thermalising the SM before `activating' 
its interaction with the TLS or 2) thermalising the SM whilst the two subsystems are interacting. The first method can be suitably described by a thermalised distribution of Fock states of the SM. The second leads to a thermal distribution with respect to the appropriately displaced oscillator states of the SM\cite{Irish2005}. For simplicity, we focus on the second case in the results discussed here.

\section{Results}\label{sec:prelim}
 
Unless otherwise stated, simulations are carried out at $T=300$~\si{\kelvin} with $J=5$~\si{\tera\hertz}, $\Omega=100$~\si{\tera\hertz}, and $\lambda=0.05$~\si{\tera\hertz}. The initial state of the system is localised on site 1 
with the corresponding displaced oscillator in a thermal state and all spectral densities have $\omega_{p}=2J$, the bare TLS frequency. These parameters have been chosen to display the off-resonant SM behaviours most 
clearly. 
 
\begin{figure}
 \centering
 \includegraphics[scale=0.3,keepaspectratio=true]{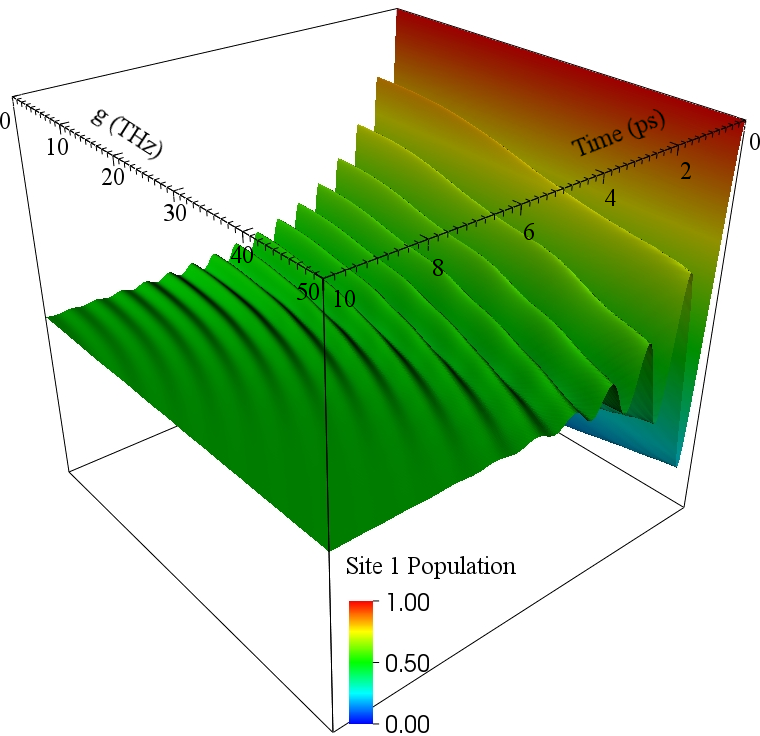}
 \caption{A 3D plot of the site 1 population dynamics as a function of $g$ for an Ohmic ($m=1$) SD. An oscillation frequency renormalisation effect is visible.}
 \label{fig:ohmic3d}
\end{figure}

We start with an Ohmic SD. The effect of the bath on the TLS, with no SM present ($g=0$), is to damp the TLS population oscillations until a thermal equilibrium steady state is reached. As can be seen in 
Fig.~\ref{fig:ohmic3d}, this characteristic remains when increasing the SM coupling strength. There is a visible damping enhancement effect brought about as $g$ becomes larger that is likely due to the mixing into higher energy 
states of the SM, which provides an increased number of decay pathways due to the broad Ohmic SD. There is also an oscillation frequency renormalisation visible as a subtle curvature in the peaks as a function of $g$. 

\begin{figure*}
\centering$
\begin{array}{cc}
\subfigure[\ m=3]{
\includegraphics[scale=0.25,keepaspectratio=true]{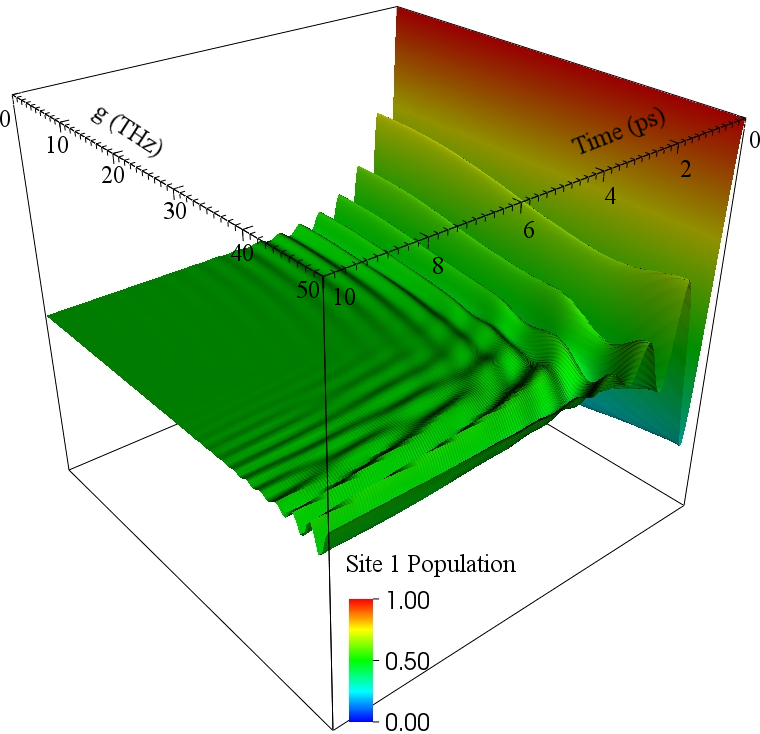}\hspace{1cm}
}&
\subfigure[\ m=5]{
\hspace{1cm}\includegraphics[scale=0.25,keepaspectratio=true]{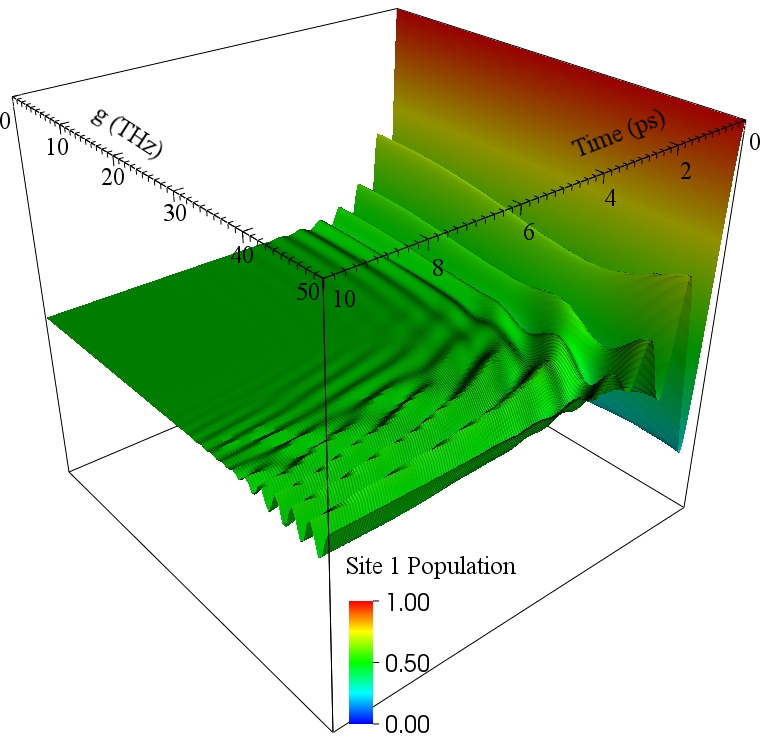}
}\vspace{0.5cm}\\*
\subfigure[\ m=7]{
\includegraphics[scale=0.25,keepaspectratio=true]{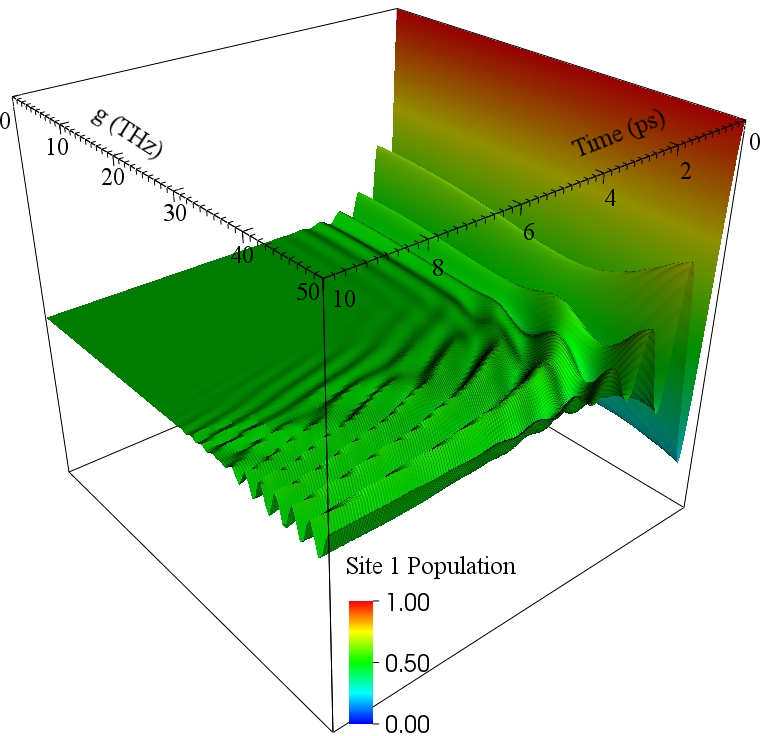}\hspace{1cm}
}&
\subfigure[\ Lorentzian]{
\hspace{1cm}\includegraphics[scale=0.25,keepaspectratio=true]{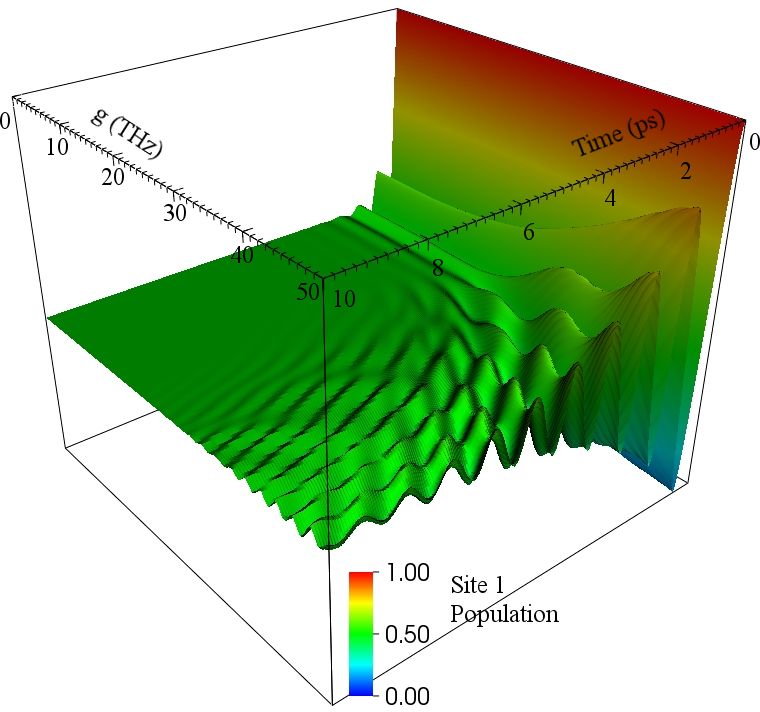}
}
\end{array}$
\caption{\label{fig:super3d} Site 1 population dynamics as a function of $g$ for super-Ohmic spectral densities with a) $m=3$, b) $m=5$, c) $m=7$ and d) Lorentzian with $W=1.5$~\si{\tera\hertz}. A clear enhancement is seen for large values of $g$, which becomes more pronounced for the more peaked spectral densities.}
\end{figure*}

To further explore these effects we proceed to display the dynamics of super-Ohmic spectral densities with $m=3,5$ and $7$, as well as the Lorentzian SD, in Fig.~\ref{fig:super3d}. The qualitative difference between the dynamics in Fig.~\ref{fig:ohmic3d} and Fig.~\ref{fig:super3d} is 
clear: as $g$ is increased the oscillations are enhanced instead of damped. The behaviour is most easily appreciated when looking in the second half of the simulated time in Fig.~\ref{fig:super3d}, where the weak-$g$ 
plateaus are replaced with population oscillations for strong~$g$. These oscillations become more pronounced with increasing $m$, as the SD becomes more peaked, and the largest effect is found for the Lorentzian SD, which is the most peaked of those displayed.

\section{Analysis and Discussion}\label{sec:ana}

Without the single mode, an initial excitation on one site of the effective TLS oscillates in time between the two sites; the amplitude of this oscillation is gradually damped by the interaction with the environment. Figs.~~\ref{fig:ohmic3d} and~\ref{fig:super3d}  show that the inclusion of the single mode complicates the dynamics considerably. However, the main features can be understood using an analytical approximation for the TLS-SM system in the absence of the bath. 

\begin{figure}
\includegraphics[scale=0.43,keepaspectratio=true]{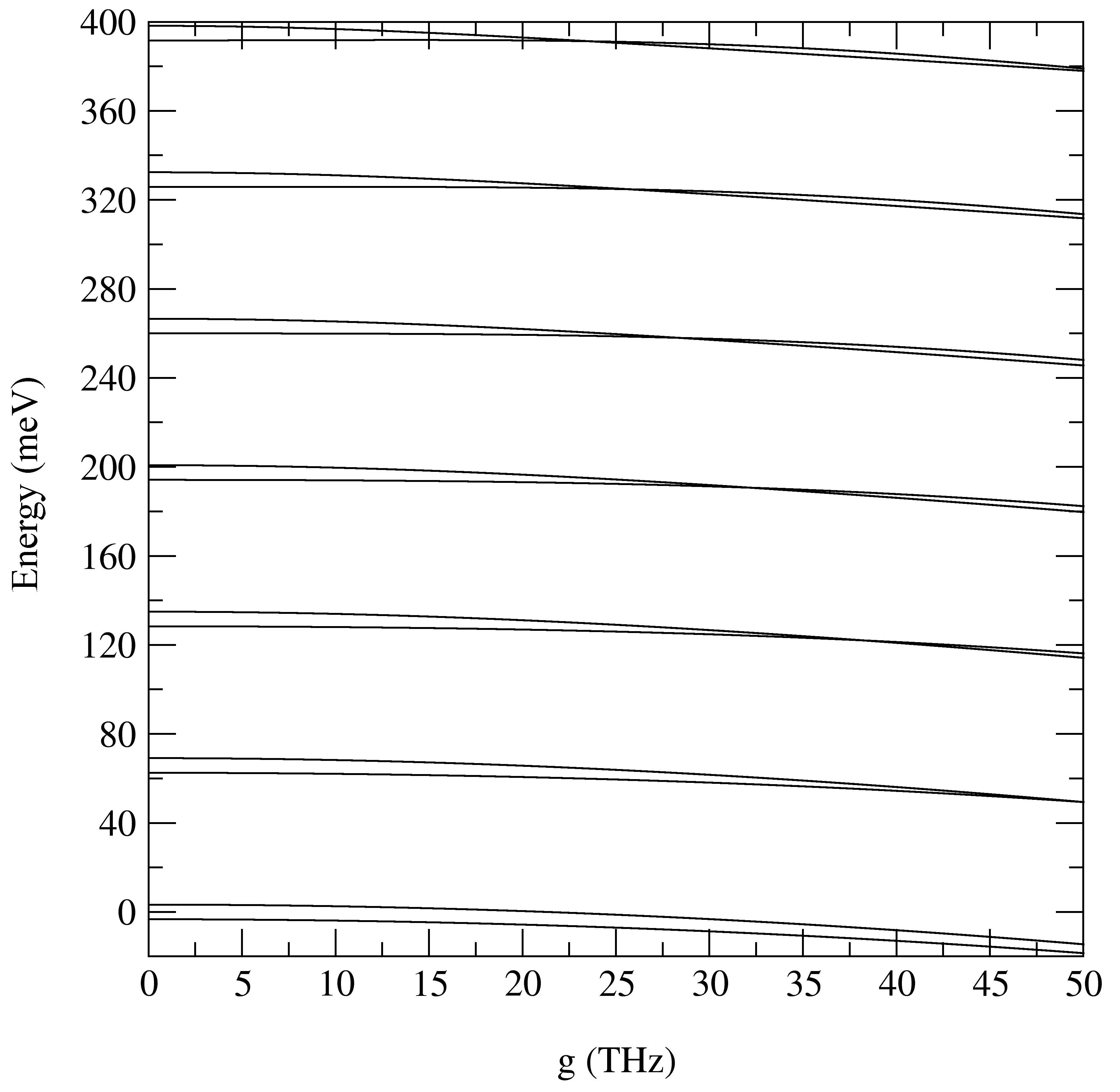}
\caption{Numerically determined energy spectrum of the TLS-SM system in the absence of the environment, as a function of the coupling $g$. The energy levels take the form of a series of closely spaced doublets.}
\label{fig:varyg} 
\end{figure}

The adiabatic approximation of Irish \textit{et al.}\cite{Irish2005} provides a physically intuitive basis for the TLS-SM system in the case, as here, where the energy splitting of the TLS (here $2J = 10$~THz) is much smaller than the frequency of the SM (here $\Omega = 100$~THz). In this parameter regime the system may be approximated by two harmonic potential wells, each associated with one site of the TLS. The effect of the interaction between the SM and the TLS is to displace each harmonic well in position space. To lowest order in the TLS energy, the eigenstates of the system are given by 
\begin{equation}
\ket{\Psi_{\pm,n}} = \frac{1}{\sqrt{2}} (\ket{0} \otimes \ket{n_0} \pm \ket{1} \otimes \ket{n_1}) ,
\end{equation}
where $\ket{i}$ corresponds to excitation of site $i$ and $\ket{n_i}$ denotes a number state of the displaced well associated with site $i$. Specifically, in terms of the original SM basis states $\ket{n}$ the displaced bases are given by $\ket{n_0} = \exp[-(g/\Omega)(\hat{a}^\dag + \hat{a})] \ket{n}$ and $\ket{n_1} = \exp[(g/\Omega)(\hat{a}^\dag + \hat{a})] \ket{n}$. The approximate energies are given by
\begin{equation}
E_{\pm,n} = \Omega \left(n - \frac{g^2}{\Omega^2} \right) \pm J e^{-2 g^2/\Omega^2} L_n\left(\frac{4g^2}{\Omega^2} \right) ,
\end{equation}
where $L_n(x)$ is a Laguerre polynomial. As shown in Fig.~\ref{fig:varyg}, the energy spectrum breaks into a series of well-spaced doublets, where each doublet corresponds to a different value of $n$.

Within this approximation, it is readily seen that an initial excitation on site $i$, with the SM in a number state in the displaced basis $\ket{n_i}$, will undergo Rabi-like oscillations between the two sites. The frequency of this oscillation is determined by the doublet splitting, $2 J e^{-2 g^2/\Omega^2} L_n(4g^2/\Omega^2)$. From this formula it is evident that one effect of increasing the coupling $g$ is a renormalisation of the oscillation frequency to smaller values, as seen in Figs.~\ref{fig:ohmic3d}~and~\ref{fig:super3d}. Moreover, the renormalised frequency depends on the number state of the SM. Hence a distribution of different $n$ values in the initial state of the SM will lead to multiple oscillation frequencies in the dynamics of the TLS population.

\begin{figure}
\centering
\subfigure[]{
\includegraphics[scale=0.35,keepaspectratio=true]{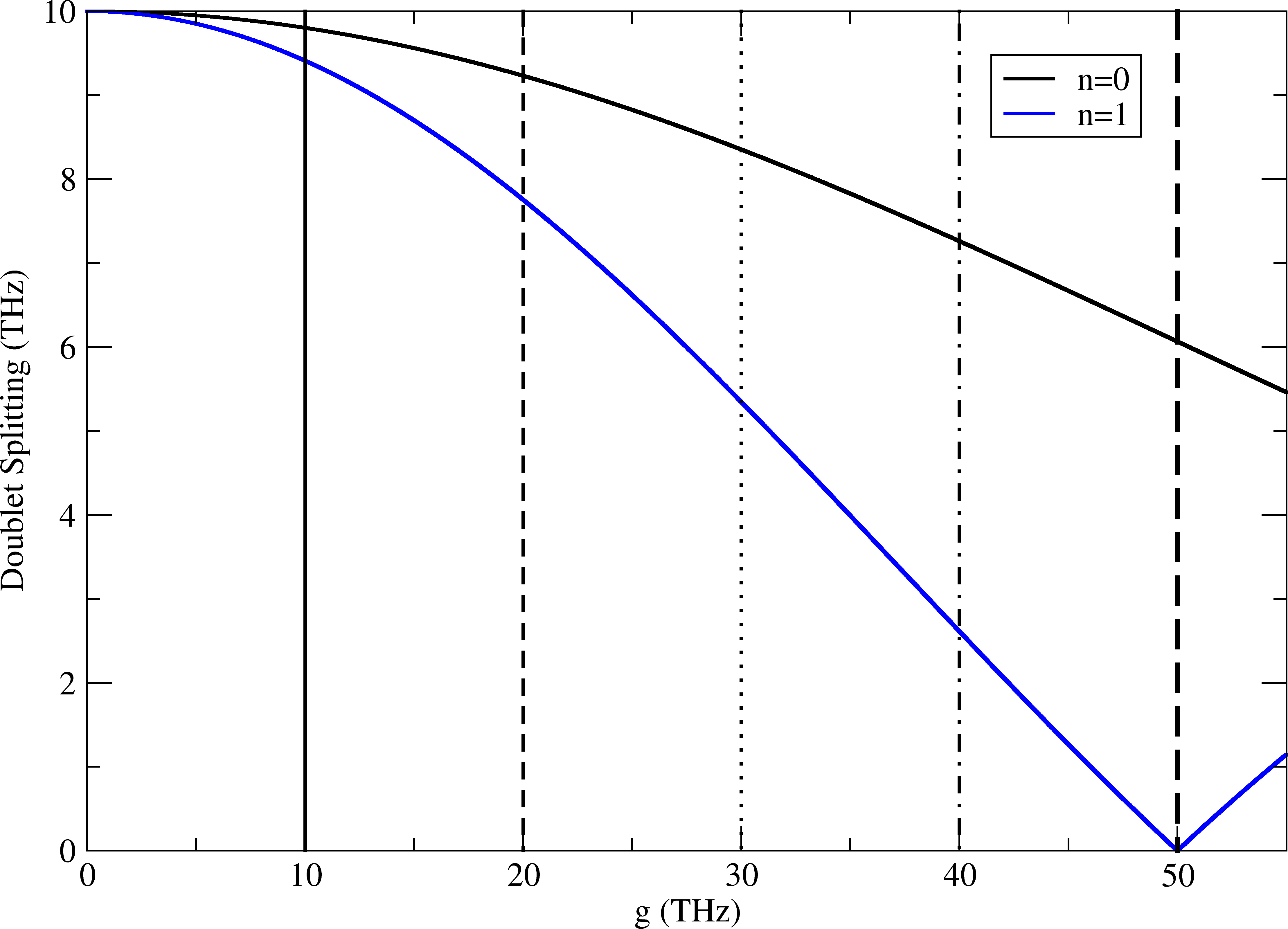}
\label{fig:splitting}}\\
\subfigure[]{
\includegraphics[scale=0.35,keepaspectratio=true]{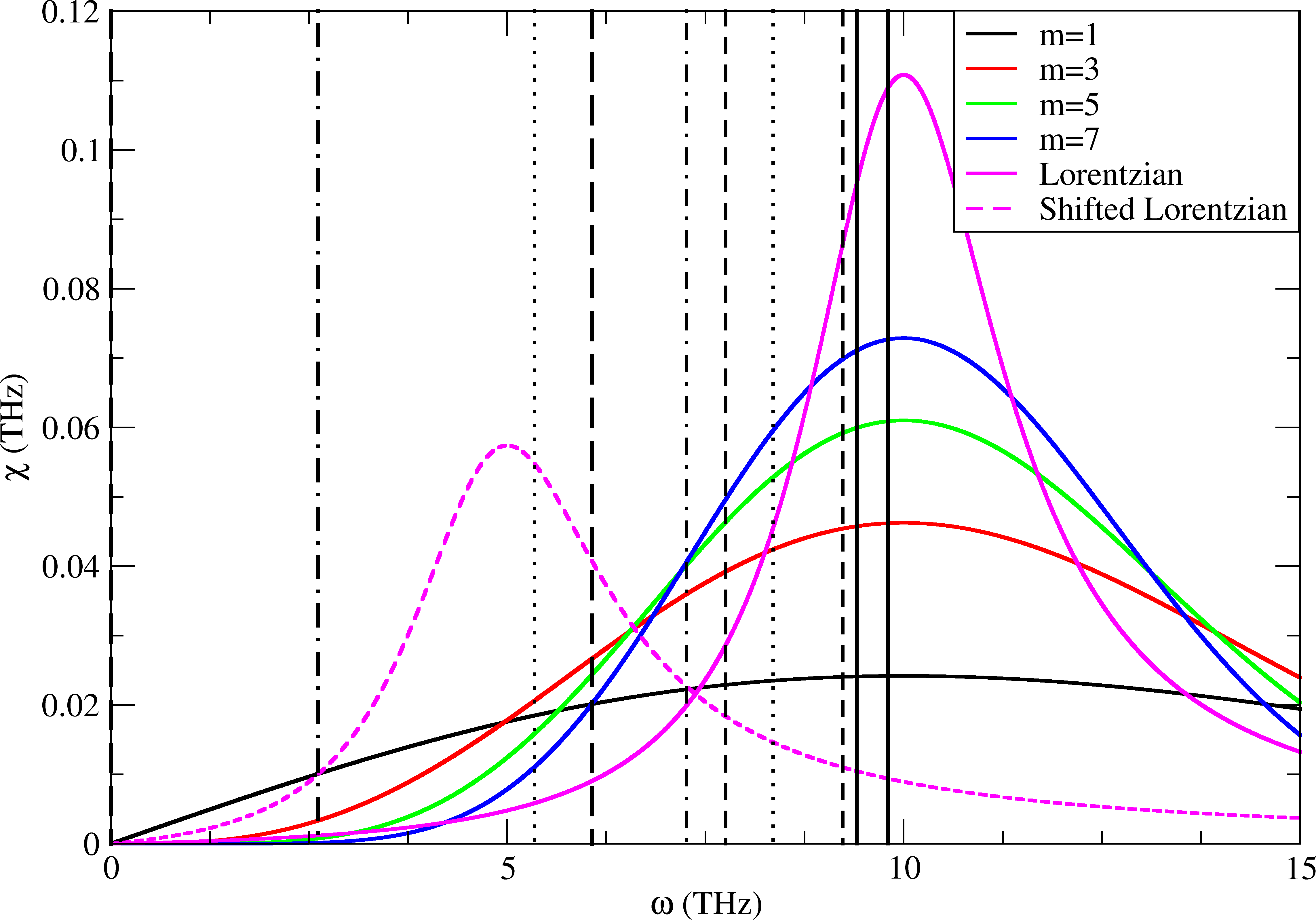}
\label{fig:sampleSD}}
\caption{(a) Energy splittings in the adiabatic approximation as a function of $g$, for the doublets corresponding to the $n=0$ (black) and $n=1$ (blue) states of the displaced SM. Vertical lines indicate the values of $g$ used in (b) below. (b) Comparison of the doublet splittings with the various spectral densities, showing the frequencies at which the spectral densities are sampled for various values of $g$. The vertical lines indicate the $n=0$ and $n=1$ splittings for the values of $g$ that have been marked in (a) with the same line style. In each case the higher frequency corresponds to $n=0$ and the lower to $n=1$. As $g$ increases, both frequencies decrease but they also move further apart.}
\end{figure}

For the SM frequency and temperature values we consider, only two doublets contribute significantly. A thermal state of a 100~THz mode at $T = 300$~K is in the ground state with probability $p(0) \approx 92\%$; the first excited state has probability $p(1) \approx 7\%$ and all other states have probability of less than 1\%. Therefore the dynamics in Figs.~\ref{fig:ohmic3d} and~\ref{fig:super3d} are dominated by just two oscillation frequencies. Fig.~\ref{fig:splitting} shows how these two frequencies change with the coupling $g$. Both frequencies display a strong renormalisation effect, shifting to smaller values as $g$ increases; however, the frequency corresponding to $n=1$ changes more rapidly with $g$.

In addition to the frequency shift seen clearly in Fig.~\ref{fig:ohmic3d}, several other aspects of the dynamics illustrated in Fig.~\ref{fig:super3d} can be explained by the frequency renormalisation effect. Fig.~\ref{fig:super3d} shows an apparently counterintuitive decrease in the damping of the oscillations as $g$ is increased. This arises from the combination of frequency renormalisation and the peaked form of the super-Ohmic spectral densities. The spectral densities have been chosen to have their peaks at the bare oscillation frequency $2J$. As $g$ increases, the oscillation frequencies shift to lower values, moving away from the maximum of the SD. Fig.~\ref{fig:sampleSD} illustrates this effect for the frequencies corresponding to the $n=0$ and $n=1$ doublets. The vertical lines indicate the renormalised frequencies for a range of $g$ values, showing the points at which the SD is sampled.
The smaller value of $\chi(\omega)$ at lower frequencies means that the system undergoes less damping. In particular, the $n=1$ oscillation frequency decreases more rapidly with $g$ and thus is damped much less than the $n=0$ mode. 

In Fig.~\ref{fig:super3d}(a)-(c) the oscillations that persist out to long times for large $g$ correspond to the $n=1$ component of the initial state, as the dominant $n=0$ component is damped away fairly rapidly. As the SD becomes more strongly peaked, this effect becomes more pronounced, which is consistent with the trend visible in Fig.~\ref{fig:super3d}. The oscillations along the $g$-axis arise from the $g$-dependent shift in frequency of the $n=1$ component. By contrast, the dominant oscillation visible in Fig.~\ref{fig:super3d}(d) at large $g$ values corresponds to the $n=0$ renormalised frequency. As seen in Fig.~\ref{fig:sampleSD}, the Lorentzian SD at this frequency is much smaller than the Ohmic or super-Ohmic SDs and hence the $n=0$ oscillation decays much more slowly.

\begin{figure}
\centering
\includegraphics[scale=0.4,keepaspectratio=true]{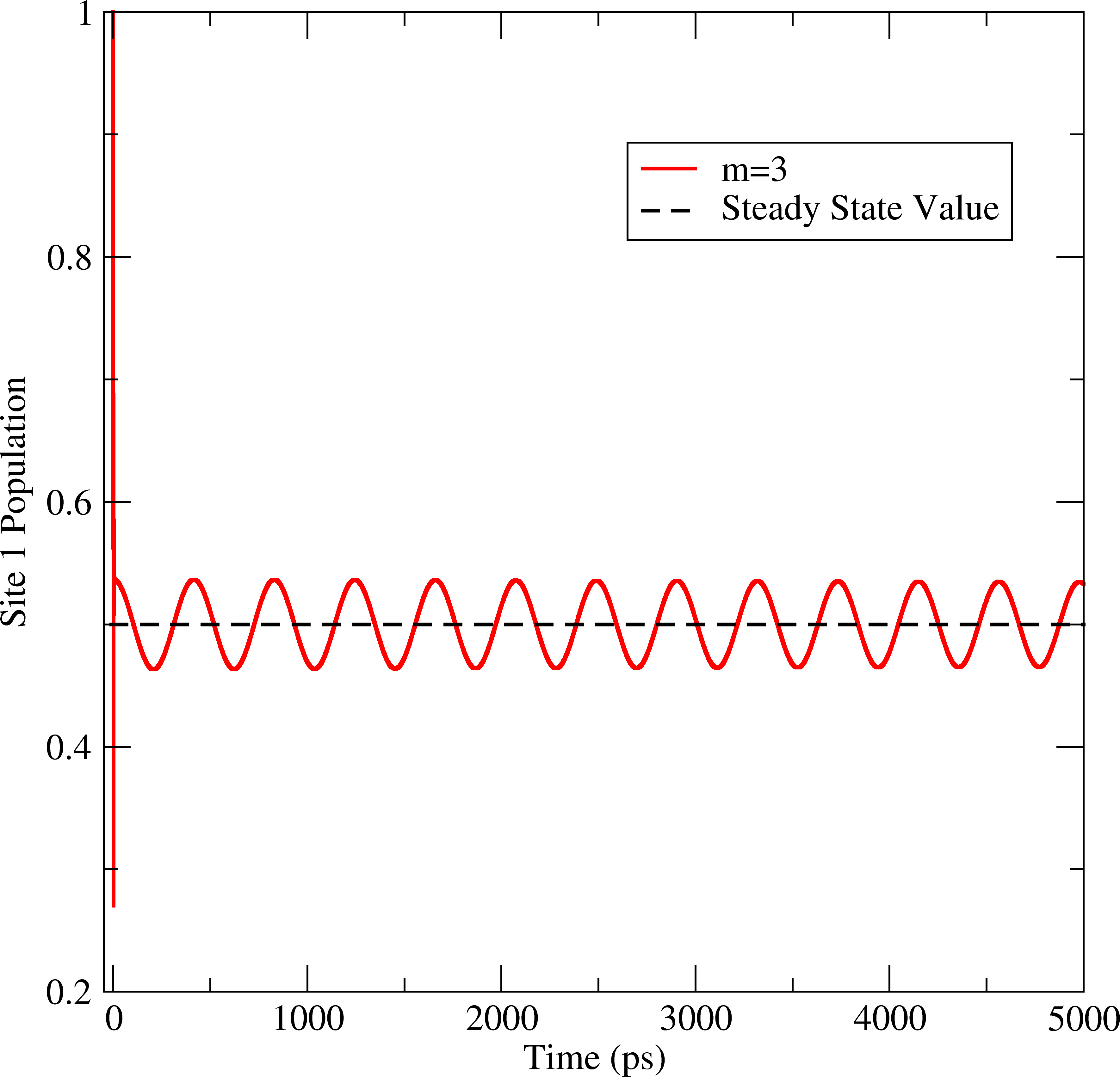}
\caption{An example of long time dynamics for a strongly coupled TLS-SM system, $g=50$~\si{\tera\hertz}. The frequency and amplitude of the oscillation correspond to the $n=1$ state of the SM. Strong frequency renormalisation due to the large value of $g$ means that the bath has an extremely weak damping effect, allowing coherent oscillations to persist out to very long times despite the influence of the environment.}
\label{fig:longdyn}
\end{figure}

In Fig.~\ref{fig:longdyn} we show long-time, nanosecond dynamics for the $m=3$ super-Ohmic SD and a very strong coupling of $g=50$~THz. There is a very long-lived oscillatory component to the dynamics that shows no signs of decay even after $5$~\si{\nano\second} -- though we have checked numerically that it does eventually reach a steady state population of 0.5.
This feature is also visible, but not so obvious, in Fig.~\ref{fig:super3d}(a)-(c) through the seemingly fixed population, away from the equilibrium population of $0.5$, for $t\gtrsim6$~\si{\pico\second}. An initial transient regime exists over a period of tens of \si{\pico\second} and can just be discerned in Fig.~\ref{fig:longdyn}. During this transient stage other oscillatory frequencies present in the dynamics are damped away leaving only the single slow feature associated with the $n=1$ doublet which, as can be seen in Fig.~\ref{fig:splitting}, is close to degeneracy at $g=50$~THz. The amplitude of this persistent slow oscillation corresponds to the probability of the $n=1$ state in the initial thermal distribution of the SM, here about 7\%.

\begin{figure}
 \centering
 \includegraphics[scale=0.3,keepaspectratio=true]{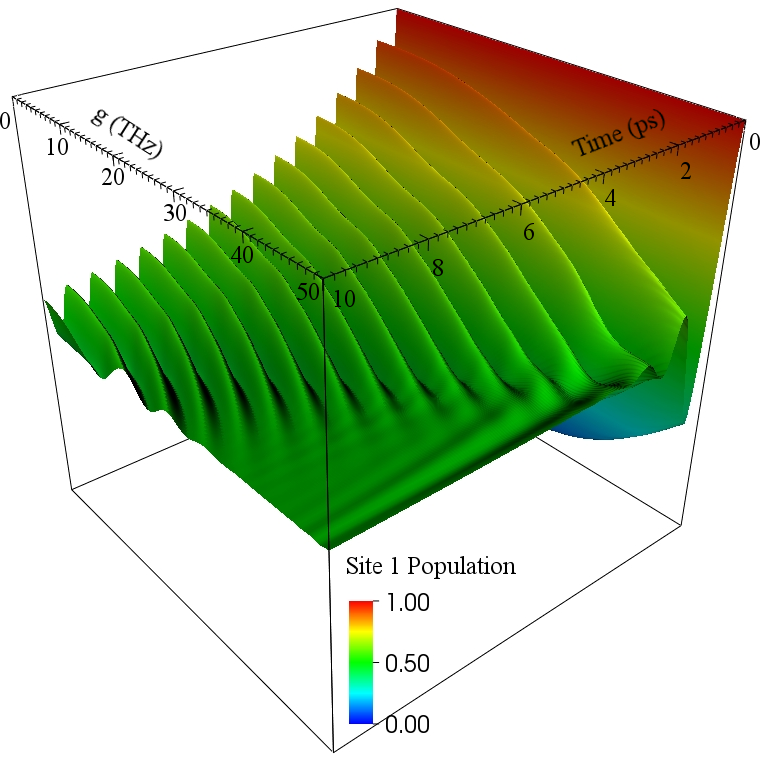}
 \caption{Site 1 population dynamics of the TLS where the bath SD is given by a Lorentzian with peak frequency $\omega_p = J = 5$~THz. In this case increasing $g$ pushes the oscillation frequencies of the system toward the peak of the SD rather than away from it, resulting in greater damping.}
 \label{fig:push}
\end{figure}

This interpretation of our results predicts that the effect of increasing $g$ should be reversed if the peak of the SD lies at a frequency below the bare $2J$ splitting of the TLS. Then as $g$ increases the renormalised oscillation frequencies will be pushed toward the peak of the SD rather than away from it, resulting in increased damping. In Fig.~\ref{fig:push} we present dynamics for such a scenario, where the peak of the Lorentzian SD is chosen to be $\omega_{p}=J=5$~\si{\tera\hertz}, shown as the shifted Lorentzian in Fig.~\ref{fig:sampleSD}. Note that the choice to keep the renormalisation energy $\lambda$ constant means that the peak height is lower than the Lorentzian centred at $\omega_{p} = 2J$. For values of $g \sim 40-50$~THz, the renormalised $n=0$ frequency corresponds to a point where $\chi_L(\omega)$ is increasing rapidly; the corresponding increase in damping is dramatically evident in the dynamics. 


\section{Conclusion}\label{sec:conc}
We have found that a single strongly coupled bosonic mode can have a profound effect on the coherent dynamics of an open TLS. Contrary to naive expectations, our work shows that strong coupling to a vibrational mode can actually 
enhance quantum coherent features in site-to-site dynamics. Moreover we have shown that this effect happens for a mode that is far off-resonant with the TLS's natural oscillation frequency and that it can occur at room 
temperature. We have also found that certain spectral densities exhibit this coherence enhancement to differing degrees; this is a consequence of their peakedness, and so how quickly the frequencies at which the SD is sampled are moved away from the peak by the SM coupling. 

 
Our results show that off-resonant modes contribute to the dynamics when treated exactly, providing justification for efforts extending theoretical tools to accommodate such treatments. If an open TLS interacts with an 
environment with a SD like those investigated here we have shown that its coherence properties are altered. 
SD modification has recently been demonstrated through reservoir engineering\cite{Haeberlein2015}, and we speculate that using this technology with a strongly coupled vibrational mode could lead to a tool for maintaining or quickly damping oscillatory dynamics.
Future work will need to generalise our approach to explore asymmetric dimers strongly coupled to more than one mode. Nonetheless, our findings provide a tantalising look at the possibilities for novel quantum effects in complex open quantum systems.

\acknowledgments

We would like to thank Ahsan Nazir and Erik Gauger for useful discussions. This work was supported by the Leverhulme Trust (RPG-080) and the EPSRC (EP/G03673X/1).


%


\appendix{}
\section{Fulton-Gouterman Transformation}\label{app:FGT}
The way that the FGT isolates parity subspaces can be seen if we take the Hamiltonian in Eq.~\ref{eq:singleSBM} and transform it from the site into the superposition basis, 
$|\pm\rangle=\frac{1}{\sqrt{2}}\bigl(|0\rangle\pm|1\rangle\bigr)$. Writing out the matrix representation of the resulting Hamiltonian (keeping the SM in the Fock basis) we get
\begin{widetext}
\begin{equation}
H=\bordermatrix{~ 	& |+0\rangle & |-0\rangle & |+1\rangle & |-1\rangle & \cdots \cr
		  \langle+0| & -J & 0 & 0 & -h & \cdots \cr
		  \langle-0| & 0 & J & -h & 0 & \cdots \cr
		  \langle+1| & 0 & -h & -J+\Omega & 0 & \cdots \cr
		  \langle-1| & -h & 0 & 0 & J+\Omega & \cdots \cr
		  \vdots & \vdots & \vdots & \vdots & \vdots & \ddots \cr}.
\end{equation}
If we look at even and odd excitation parity states we see only elements linking states of like-parity meaning one can build two subspace Hamiltonians,
\begin{equation}
H^{+}=\bordermatrix{~ 	& |+0\rangle & |-1\rangle & \cdots \cr
		  \langle+0| & -J & -h & \cdots \cr
		  \langle-1| &-h & J+\Omega & \cdots \cr
		  \vdots & \vdots & \vdots & \ddots \cr},\quad
H^{-}=\bordermatrix{~ 	& |-0\rangle & |+1\rangle & \cdots \cr
		  \langle-0| & J & -h & \cdots \cr
		  \langle+1| & -h & -J+\Omega & \cdots \cr
		  \vdots & \vdots & \vdots & \ddots \cr}.
\end{equation}
\end{widetext}
In operator form these subspace Hamiltonians are the same as those in Eq.~\ref{eq:FGTsubs} which are obtained after direct application of the FGT.
\section{Interaction Picture}\label{app:interact}
 
The interaction picture form of our interaction Hamiltonian, $H_{I}(t)$, is reached using a unitary operator approach,
\begin{equation}
H_{I}(t)=\mathrm{e}^{i(H_{\rm S}+H _{\rm B})t}H_{I}\mathrm{e}^{-i(H_{\rm S}+H _{\rm B})t},
\end{equation}
where all of the Hamiltonians on the RHS are in their Schr\"{o}dinger form as found in Eq.~\ref{eq:hamil}, with $H_B = \sum_{\mathbf{q}}\omega_{\mathbf{q}}\hat{n}_{\mathbf{q}}$ and 
$H_I = \hat{\sigma}_{x}\sum_{\mathbf{q}}h_{\mathbf{q}}(\hat{b}_{\mathbf{q}}^{\dagger}+\hat{b}_{\mathbf{q}})$. Since $[H_{\rm S},H _{\rm B}]=0$ and since $H_{I}$ can be written in a separable form we can also formulate 
$H_{I}(t)$ in a separable form:
\begin{equation}\label{eq:UniIntSep}
H_{I}(t)=\mathrm{e}^{i H_{\rm S}t}\hat{\sigma}_{x}\mathrm{e}^{-i H_{\rm S}t}\otimes\mathrm{e}^{i H _{\rm B}t}\displaystyle\sum_{\mathbf{q}}h_{\mathbf{q}}\bigl(\hat{b}_{\mathbf{q}}^{\dagger}+\hat{b}_{\mathbf{q}}\bigr)\mathrm{e}^{-i H _{\rm B}t}.
\end{equation}
Now the benefit of expressing $\sigma_{x}$ in terms of the FGT eigenstates, $\{|\psi^{\pm}_{k}\rangle\}$, can be seen: the exponentiated $H_{\rm S}$ will simply act on its eigenstates in Eq.~\ref{eq:UniIntSep} and be replaced 
with the corresponding eigenvalues (which can be found numerically). 
 
To change its basis we first split $\sigma_{x}$ into the constituent parts $\sigma_{+}=|+\rangle\langle-|$ and $\sigma_{-}=|-\rangle\langle+|$, the TLS raising and lowering operators ($\sigma_{x}=\sigma_{+}+\sigma_{-}$). The 
system eigenstates in Eq.~\ref{eq:FGstates} exist in a Hilbert space spanning the TLS and the SM so we need 
\begin{equation}
\sigma_{+n}=\sigma_{+}\otimes I=\displaystyle\sum_{n}|+n\rangle\langle-n| 
\end{equation}
and the corresponding equation for $\sigma_{-n}$. Next we expand the eigenstates in the superposition basis: Eq.~\ref{eq:FGstates} becomes
\begin{equation}\label{eq:midFGstates}
|\psi_{k}^{\pm}\rangle=\frac{1}{2}\Bigl[\bigl(\hat{P}\pm1\bigr)|+\rangle-\bigl(\hat{P}\mp1\bigr)|-\rangle\Bigr]|\phi^{\pm}_{k}\rangle.
\end{equation}
The oscillator subspace eigenstates are expressed in a Fock basis as 
\begin{equation}\label{eq:FockSubExp}
|\phi^{\pm}_{k}\rangle=\displaystyle\sum_{n}C^{\pm}_{kn}|n\rangle
\end{equation}
with the coefficients, $C^{\pm}_{kn}$, obtained after numerical solution of the eigensystem via $C^{\pm}_{kn}=\langle n|\phi_{k}^{\pm}\rangle$. With these definitions we can see that the $|\pm\rangle$ states pick out 
different parity Fock states due to their relation with $\hat{P}$ in Eq.~\ref{eq:midFGstates}; our eigenstates become
\begin{equation}
|\psi_{k}^{\pm}\rangle=\displaystyle\sum_{even\, n}C^{\pm}_{kn}|\pm n\rangle+\displaystyle\sum_{odd\, n}C^{\pm}_{kn}|\mp n\rangle.
\end{equation}
In this form it can clearly be seen that $\langle\psi^{\pm}_{k}|\psi^{\mp}_{k'}\rangle=0$ due to the orthogonal natures of the Fock and superposition states.
 
The change of basis of the TLS raising and lowering operators is done by calculating the overlap of the eigenstates with the operators in order to find the corresponding matrix elements:
\begin{align}
\langle\psi^{\pm}_{k}|\sigma_{+n}|\psi^{\pm}_{k'}\rangle & =\langle\psi^{\pm}_{k}|\sigma_{-n}|\psi^{\pm}_{k'}\rangle=0 \\
\langle\psi^{\pm}_{k}|\sigma_{\pm n}|\psi^{\mp}_{k'}\rangle & =\displaystyle\sum_{even\,n}C^{\pm}_{nk}C^{\mp}_{k'n} \\
\langle\psi^{\mp}_{k}|\sigma_{\pm n}|\psi^{\pm}_{k'}\rangle & =\displaystyle\sum_{odd\,n}C^{\mp}_{nk}C^{\pm}_{k'n}.
\end{align}
The reversed subscripts on the first $C^{\pm}$ in each pair (compared with the definition in Eq.~\ref{eq:FockSubExp}) is to be interpreted as complex conjugation, $C^{\pm}_{nk}=\bigl(C^{\pm}_{kn}\bigr)^{*}$. Finally our 
raising and lowering operators are
\begin{widetext}
\begin{equation}\label{eq:pmOps}
\sigma_{\pm n}=\displaystyle\sum_{k,k'}\biggl\{\displaystyle\sum_{even\,n}C^{\pm}_{nk}C^{\mp}_{k'n}|\psi_{k}^{\pm}\rangle\langle\psi_{k'}^{\mp}|+\displaystyle\sum_{odd\,n}C^{\mp}_{nk}C^{\pm}_{k'n}|\psi_{k}^{\mp}\rangle\langle\psi_{k'}^{\pm}|\biggr\}.
\end{equation}
Building the $\sigma_{xn}$ operator removes one layer of complexity from the operators in Eq.~\ref{eq:pmOps}; adding them together means the even $n$ terms are added to the odd $n$ terms leaving a sum over all $n$,
\begin{equation}\label{eq:FGX}
\sigma_{x n}=\displaystyle\sum_{k,k',n}\biggl\{C^{+}_{nk}C^{-}_{k'n}|\psi_{k}^{+}\rangle\langle\psi_{k'}^{-}|+C^{-}_{nk}C^{+}_{k'n}|\psi_{k}^{-}\rangle\langle\psi_{k'}^{+}|\biggr\}.
\end{equation}
We can now bring $\sigma_{x n}$ into the interaction picture using Eq.~\ref{eq:UniIntSep} giving
\begin{equation}
\mathrm{e}^{i H_{0}^{S}t}\hat{\sigma}_{xn}\mathrm{e}^{-i H_{0}^{S}t}=\displaystyle\sum_{k,k',n}\biggl\{C^{+}_{nk}C^{-}_{k'n}\mathrm{e}^{i\Lambda_{kk'}t}|\psi_{k}^{+}\rangle\langle\psi_{k'}^{-}|+C^{-}_{nk'}C^{+}_{kn}\mathrm{e}^{-i\Lambda_{kk'}t}|\psi_{k'}^{-}\rangle\langle\psi_{k}^{+}|\biggr\}.
\end{equation}
\end{widetext}
Here the indices on the second element have been exchanged to allow for a common constant, $\Lambda_{kk'}$, to exist. This is comprised of the eigenvalues of the two states contributing to the element,
\begin{equation}\label{eq:lambdas}
\Lambda_{kk'}=E^{+}_{k}-E^{-}_{k'}.
\end{equation}
 
The bath operator part of the separable Eq.~\ref{eq:UniIntSep} moves into the interaction picture using the identities
\begin{align}
\mathrm{e}^{\alpha\hat{n}}\hat{b}\mathrm{e}^{-\alpha\hat{n}} & =\hat{b}\mathrm{e}^{-\alpha} \\
\mathrm{e}^{\alpha\hat{n}}\hat{b}^{\dagger}\mathrm{e}^{-\alpha\hat{n}} & =\hat{b}^{\dagger}\mathrm{e}^{\alpha}. 
\end{align}
This allows us to compute Eq.~\ref{eq:UniIntSep} and express the full interaction picture, interaction Hamiltonian as
\begin{widetext}
\begin{equation}
H_{I}^{I}(t)=\displaystyle\sum_{k,k'}\mathrm{e}^{i\Lambda_{kk'}t}\hat{B}^{\dagger}(t)\hat{\zeta}^{+}_{kk'}+\mathrm{e}^{-i\Lambda_{kk'}t}\hat{B}^{\dagger}(t)\hat{\zeta}^{-}_{k'k}+\mathrm{e}^{i\Lambda_{kk'}t}\hat{B}(t)\hat{\zeta}^{+}_{kk'}+\mathrm{e}^{-i\Lambda_{kk'}t}\hat{B}(t)\hat{\zeta}^{-}_{k'k}.
\end{equation}
\end{widetext}
The $\hat{B}(t)$ are the interaction picture bath operators such that 
\begin{equation}
\hat{B}(t)=\displaystyle\sum_{\mathbf{q}}h_{\mathbf{q}}\hat{b}_{\mathbf{q}}\mathrm{e}^{-i\omega_{\mathbf{q}}t}. 
\end{equation}
The $\hat{\zeta}$'s are Fulton-Gouterman state switching operators defined as
\begin{align}
\hat{\zeta}^{+}_{kk'} & =\displaystyle\sum_{n}C^{+}_{nk}C^{-}_{k'n}|\psi_{k}^{+}\rangle\langle\psi_{k'}^{-}| \label{eq:fgtswitch1},\\
\hat{\zeta}^{-}_{k'k} & =\displaystyle\sum_{n}C^{-}_{nk'}C^{+}_{kn}|\psi_{k'}^{-}\rangle\langle\psi_{k}^{+}| \label{eq:fgtswitch2}.
\end{align}

\section{Master Equation}\label{app:maseq}
 
Now we can proceed to evaluate the double commutator in Eq.~\ref{eq:MEgen} which requires combining the various operators we've discussed up until now; noting 
$\hat{\zeta}^{+}_{kk'}\hat{\zeta}^{+}_{ll'}=\hat{\zeta}^{-}_{k'k}\hat{\zeta}^{-}_{l'l}=0$ helps to reduce the number of terms. Due to the separable nature of the formulae presented, the partial trace over the double 
commutator boils down to a trace over the bath operators in each additive term in the expanded commutators. This again allows for a further reduction of terms as 
$\langle\hat{B}(t)\hat{B}(t')\rangle=\langle\hat{B}^{\dagger}(t)\hat{B}^{\dagger}(t')\rangle=0$, where the use of angular brackets denotes the trace over a thermal density matrix (the assumed state of the environmental 
bath). A multi-mode thermal density matrix takes the form
\begin{equation}
\rho _{\rm B}=N\displaystyle\prod_{\mathbf{p}}\mathrm{e}^{-\beta\omega_{\mathbf{p}}\hat{n}_{\mathbf{p}}},
\end{equation}
with the normalisation constant
\begin{equation}\label{eq:norm}
N =\displaystyle\prod_{\mathbf{p'}}1-\mathrm{e}^{-\beta\omega_{\mathbf{p'}}}.
\end{equation}
This could also be written as $N=\prod_{i}N_{i}$ where $N_{i}=1-\mathrm{e}^{-\beta\omega_{i}}$.

Now let us evaluate the non-zero bath correlation functions $\langle\hat{B}^{\dagger}(t)\hat{B}(t')\rangle$ and $\langle\hat{B}(t)\hat{B}^{\dagger}(t')\rangle$. We show how to explicitly evaluate the first of these:
\begin{widetext}
\begin{align}\label{eq:BathTrace1}
\langle\hat{B}^{\dagger}(t)\hat{B}(t')\rangle & =N\displaystyle\sum_{\{n\}}\langle n_{0},n_{1}...|\displaystyle\sum_{\mathbf{q}}h_{\mathbf{q}}\hat{a}^{\dagger}_{\mathbf{q}}\mathrm{e}^{i\omega_{\mathbf{q}}t}\displaystyle\sum_{\mathbf{q'}}h_{\mathbf{q'}}\hat{a}_{\mathbf{q'}}\mathrm{e}^{-i\omega_{\mathbf{q'}}t'}\displaystyle\prod_{\mathbf{p}}\mathrm{e}^{-\beta\omega_{\mathbf{p}}\hat{n}_{\mathbf{p}}}|n_{0},n_{1}...\rangle \nonumber \\
& =N\displaystyle\sum_{\{n\}}\langle n_{0},n_{1}...|\displaystyle\sum_{\mathbf{q}}h^{2}_{\mathbf{q}}\hat{n}_{\mathbf{q}}\mathrm{e}^{i\omega_{\mathbf{q}}(t-t')}\displaystyle\prod_{\mathbf{p}}\mathrm{e}^{-\beta\omega_{\mathbf{p}}\hat{n}_{\mathbf{p}}}|n_{0},n_{1}...\rangle \nonumber \\
& =N\displaystyle\sum_{\{n\}}\langle n_{0},n_{1}...|h^{2}_{0}\hat{n}_{0}\mathrm{e}^{i\omega_{0}(t-t')}\displaystyle\prod_{\mathbf{p}}\mathrm{e}^{-\beta\omega_{\mathbf{p}}\hat{n}_{\mathbf{p}}}+h^{2}_{1}\hat{n}_{1}\mathrm{e}^{i\omega_{1}(t-t')}\displaystyle\prod_{\mathbf{p}}\mathrm{e}^{-\beta\omega_{\mathbf{p}}\hat{n}_{\mathbf{p}}}+...|n_{0},n_{1}...\rangle \nonumber \\
& =N\displaystyle\sum_{\{n\}}\biggl[\langle n_{0},n_{1}...|h^{2}_{0}\hat{n}_{0}\mathrm{e}^{i\omega_{0}(t-t')}\displaystyle\prod_{\mathbf{p}}\mathrm{e}^{-\beta\omega_{\mathbf{p}}\hat{n}_{\mathbf{p}}}|n_{0},n_{1}...\rangle+\langle n_{0},n_{1}...|h^{2}_{1}\hat{n}_{1}\mathrm{e}^{i\omega_{1}(t-t')}\displaystyle\prod_{\mathbf{p}}\mathrm{e}^{-\beta\omega_{\mathbf{p}}\hat{n}_{\mathbf{p}}}|n_{0},n_{1}...\rangle+...\biggr] \nonumber \\
& =\displaystyle\prod_{i}N_{i}\displaystyle\sum_{\{n\}}\biggl[h^{2}_{0}\hat{n}_{0}\mathrm{e}^{i\omega_{0}(t-t')}\displaystyle\prod_{\mathbf{p}}\mathrm{e}^{-\beta\omega_{\mathbf{p}}\hat{n}_{\mathbf{p}}}+h^{2}_{1}\hat{n}_{1}\mathrm{e}^{i\omega_{1}(t-t')}\displaystyle\prod_{\mathbf{p}}\mathrm{e}^{-\beta\omega_{\mathbf{p}}\hat{n}_{\mathbf{p}}}+...\biggr].
\end{align}
\end{widetext}
In the last line of Eq.~\ref{eq:BathTrace1} each instance of the product over $\mathbf{p}$ has all of its elements cancel with the elements in the normalisation product (Eq.~\ref{eq:norm}), except when $\mathbf{p}$ equals 
the index of the term the product is a part of. This leads to
\begin{align}\label{eq:BathTrace2}
\langle\hat{B}^{\dagger}(t)\hat{B}(t')\rangle & =\displaystyle\sum_{\mathbf{p}}N_{\mathbf{p}}h_{\mathbf{p}}^{2}\mathrm{e}^{i\omega_{\mathbf{p}}(t-t')}\displaystyle\sum_{n_{\mathbf{p}}}n_{\mathbf{p}}\mathrm{e}^{-\beta\omega_{\mathbf{p}}n_{\mathbf{p}}} \nonumber \\
& =\displaystyle\sum_{\mathbf{p}}h_{\mathbf{p}}^{2}\frac{\mathrm{e}^{i\omega_{\mathbf{p}}(t-t')}}{\mathrm{e}^{\beta\omega_{\mathbf{p}}}-1}.
\end{align}
 
The last step in the process of solving the partial trace involves the SD,
\begin{equation}\label{eq:specden}
\chi(\omega)=\displaystyle\sum_{\mathbf{q}}|h_{\mathbf{q}}|^{2}\,\delta(\omega-\omega_{\mathbf{q}}),
\end{equation}
which uses the bath coupling factors, $h_{\mathbf{q}}$, to describe the action of the bath. Due to the delta function summation we can write the identity
\begin{align}
\int\mathrm{d}\omega\,\chi(\omega)\phi(\omega) & =\int\mathrm{d}\omega\displaystyle\sum_{\mathbf{k}}h_{\mathbf{k}}^{2}\delta(\omega-\omega_{\mathbf{k}})\phi(\omega) \nonumber \\
& =\displaystyle\sum_{\mathbf{k}}h_{\mathbf{k}}^{2}\int\mathrm{d}\omega\,\phi(\omega)\delta(\omega-\omega_{\mathbf{k}}) \nonumber \\
& =\displaystyle\sum_{\mathbf{k}}h_{\mathbf{k}}^{2}\phi(\omega_{\mathbf{k}}).
\end{align}
Applying this to Eq.~\ref{eq:BathTrace2}, gives finally
\begin{equation}\label{eq:BathTrace3}
\langle\hat{B}^{\dagger}(t)\hat{B}(t')\rangle=\int\mathrm{d}\omega\,\chi(\omega)\frac{\mathrm{e}^{i\omega(t-t')}}{\mathrm{e}^{\beta\omega}-1}.
\end{equation}
The evolution of the second bath correlation function follows similarly, using $\hat{a}\hat{a}^{\dagger}=\hat{n}+1$ instead of $\hat{a}^{\dagger}\hat{a}=\hat{n}$. This leads to
\begin{equation}\label{eq:BathTrace4}
\langle\hat{B}(t)\hat{B}^{\dagger}(t')\rangle=\int\mathrm{d}\omega\,\chi(\omega)\frac{\mathrm{e}^{-i\omega(t-t')}}{1-\mathrm{e}^{-\beta\omega}}.
\end{equation}

 
By employing the identity
\begin{equation}\label{eq:deltaIdent}
\displaystyle\int\limits_{0}^{\infty}\mathrm{d}s\,\mathrm{e}^{i(a-b)s}=\pi\delta(a-b)+\frac{P}{i(a-b)}
\end{equation}
(where we neglect the second, principle value, term which would lead to only small Lamb shifts) the integrals over $s$, in Eq.~\ref{eq:MEgen}, and $\omega$, introduced by the bath correlation functions, can now be performed. Let us look at one of the four terms produced when expanding the double commutator present in Eq.~\ref{eq:MEgen}:
\begin{widetext}
\begin{align}
H_{I}^{I}(t)H_{I}^{I}(t-s)\rho(t)=\displaystyle\sum_{k,k',p,p'}\biggl[\langle\hat{B}^{\dagger}(t)\hat{B}(t-s)\rangle & +\langle\hat{B}(t)\hat{B}^{\dagger}(t-s)\rangle\biggr]\times \nonumber \\
				 & \biggl[\mathrm{e}^{i(\Lambda_{kk'}-\Lambda_{pp'})t}\mathrm{e}^{i\Lambda_{pp'}s}\hat{\zeta}^{+}_{kk'}\hat{\zeta}^{-}_{p'p}\rho_{s}(t)+\mathrm{e}^{-i(\Lambda_{kk'}-\Lambda_{pp'})t}\mathrm{e}^{-i\Lambda_{pp'}s}\hat{\zeta}^{-}_{k'k}\hat{\zeta}^{+}_{pp'}\rho_{s}(t)\biggr].
\end{align}
\end{widetext}
Again to proceed we shall explicitly show here one example of how to evaluate these terms with the others being similarly computable:
\begin{align}\label{eq:startgam}
\int\limits_{0}^{\infty}\mathrm{d}s\,\langle\hat{B}^{\dagger}(t)\hat{B}(t-s)\rangle\mathrm{e}^{\pm i\Lambda_{pp'}s} & =\int\limits_{0}^{\infty}\mathrm{d}s\int\limits_{0}^{\infty}\mathrm{d}\omega\,\chi(\omega)\frac{\mathrm{e}^{i(\omega\pm\Lambda_{pp'})s}}{\mathrm{e}^{\beta\omega}-1} \nonumber \\
& =\int\limits_{0}^{\infty}\mathrm{d}\omega\,\chi(\omega)\frac{\delta(\omega\pm\Lambda_{pp'})}{\mathrm{e}^{\beta\omega}-1} \nonumber \\
& =\frac{\pi\chi(\mp\Lambda_{pp'})}{\mathrm{e}^{\mp\beta\Lambda_{pp'}}-1}=\Gamma(\mp\Lambda_{pp'}).
\end{align}
We have defined $\Gamma$ (and also shortly $\Gamma'$) to simplify future equations; it describes the thermal decay rate of the associated transition denoted by $\Lambda$. The three remaining terms are:
\begin{align}
\int\limits_{0}^{\infty}\mathrm{d}s\,\langle\hat{B}^{\dagger}(t-s)\hat{B}(t)\rangle\mathrm{e}^{\pm i\Lambda_{pp'}s} & =\frac{\pi\chi(\pm\Lambda_{pp'})}{\mathrm{e}^{\pm\beta\Lambda_{pp'}}-1}=\Gamma(\pm\Lambda_{pp'}) \\
\int\limits_{0}^{\infty}\mathrm{d}s\,\langle\hat{B}(t)\hat{B}^{\dagger}(t-s)\rangle\mathrm{e}^{\pm i\Lambda_{pp'}s} & =\frac{\pi\chi(\pm\Lambda_{pp'})}{1-\mathrm{e}^{\mp\beta\Lambda_{pp'}}}=\Gamma'(\pm\Lambda_{pp'}) \\
\int\limits_{0}^{\infty}\mathrm{d}s\,\langle\hat{B}(t-s)\hat{B}^{\dagger}(t)\rangle\mathrm{e}^{\pm i\Lambda_{pp'}s} & =\frac{\pi\chi(\mp\Lambda_{pp'})}{1-\mathrm{e}^{\pm\beta\Lambda_{pp'}}}=\Gamma'(\mp\Lambda_{pp'}). \label{eq:endgam}
\end{align}
 
We are now in a position to write out the first form of the differential equation we have been working towards from Eq.~\ref{eq:MEgen}:
\begin{widetext}
\begin{align}\label{eq:nonSA}
\frac{\mathrm{d}\rho_{s}(t)}{\mathrm{d}t}=-\displaystyle\sum_{k,k',p,p'}\biggl[ & \Bigl(\Gamma(-\Lambda_{pp'})+\Gamma'(\Lambda_{pp'})\Bigr)\mathrm{e}^{i(\Lambda_{kk'}-\Lambda_{pp'})t}\hat{\zeta}^{+}_{kk'}\hat{\zeta}^{-}_{p'p}\rho_{s}(t) \nonumber \\
+ & \Bigl(\Gamma(\Lambda_{pp'})+\Gamma'(-\Lambda_{pp'})\Bigr)\mathrm{e}^{-i(\Lambda_{kk'}-\Lambda_{pp'})t}\hat{\zeta}^{-}_{k'k}\hat{\zeta}^{+}_{pp'}\rho_{s}(t) \nonumber \\
+ & \Bigl(\Gamma(-\Lambda_{kk'})+\Gamma'(\Lambda_{kk'})\Bigr)\mathrm{e}^{i(\Lambda_{kk'}-\Lambda_{pp'})t}\rho_{s}(t)\hat{\zeta}^{+}_{kk'}\hat{\zeta}^{-}_{p'p} \nonumber \\
+ & \Bigl(\Gamma(\Lambda_{kk'})+\Gamma'(-\Lambda_{kk'})\Bigr)\mathrm{e}^{-i(\Lambda_{kk'}-\Lambda_{pp'})t}\rho_{s}(t)\hat{\zeta}^{-}_{k'k}\hat{\zeta}^{+}_{pp'} \nonumber \\ 
- & \Bigl(\Gamma(-\Lambda_{pp'})+\Gamma'(\Lambda_{pp'})\Bigr)\Bigl(\mathrm{e}^{i(\Lambda_{kk'}+\Lambda_{pp'})t}\hat{\zeta}^{+}_{kk'}\rho_{s}(t)\hat{\zeta}^{+}_{pp'}+\mathrm{e}^{-i(\Lambda_{kk'}-\Lambda_{pp'})t}\hat{\zeta}^{-}_{k'k}\rho_{s}(t)\hat{\zeta}^{+}_{pp'}\Bigr) \nonumber \\
- & \Bigl(\Gamma(\Lambda_{pp'})+\Gamma'(-\Lambda_{pp'})\Bigr)\Bigl(\mathrm{e}^{-i(\Lambda_{kk'}+\Lambda_{pp'})t}\hat{\zeta}^{-}_{k'k}\rho_{s}(t)\hat{\zeta}^{-}_{p'p}+\mathrm{e}^{i(\Lambda_{kk'}-\Lambda_{pp'})t}\hat{\zeta}^{+}_{kk'}\rho_{s}(t)\hat{\zeta}^{-}_{p'p}\Bigr) \nonumber \\
- & \Bigl(\Gamma(\Lambda_{kk'})+\Gamma'(-\Lambda_{kk'})\Bigr)\Bigl(\mathrm{e}^{i(\Lambda_{kk'}+\Lambda_{pp'})t}\hat{\zeta}^{+}_{kk'}\rho_{s}(t)\hat{\zeta}^{+}_{pp'}+\mathrm{e}^{i(\Lambda_{kk'}-\Lambda_{pp'})t}\hat{\zeta}^{+}_{kk'}\rho_{s}(t)\hat{\zeta}^{-}_{p'p}\Bigr) \nonumber \\
- & \Bigl(\Gamma(-\Lambda_{kk'})+\Gamma'(\Lambda_{kk'})\Bigr)\Bigl(\mathrm{e}^{-i(\Lambda_{kk'}+\Lambda_{pp'})t}\hat{\zeta}^{-}_{k'k}\rho_{s}(t)\hat{\zeta}^{-}_{p'p}+\mathrm{e}^{-i(\Lambda_{kk'}-\Lambda_{pp'})t}\hat{\zeta}^{-}_{k'k}\rho_{s}(t)\hat{\zeta}^{+}_{pp'}\Bigr)\biggr].
\end{align}
\end{widetext}

In order to simplify Eq.~\ref{eq:nonSA} we employ the secular approximation which assumes that if the oscillatory exponential factors have a non-zero frequency (i.e. $\Lambda_{kk'}-\Lambda_{pp'}\neq0\neq\Lambda_{kk'}+\Lambda_{pp'}$) then they correspond to a rapid oscillation compared to the time scale over which we have assumed the system density matrix changes. These terms are 
neglected. To ensure $\Lambda_{kk'}-\Lambda_{pp'}=0$ we can insert $\delta_{k,q}\delta_{k',q'}$ into the summation. Now $\Lambda_{kk'}+\Lambda_{pp'}\rightarrow2\Lambda_{kk'}$ which in general is non-zero so these terms are neglected also (the special case of $2\Lambda_{kk}=0$ would similarly mean $\Gamma(\pm\Lambda_{kk})=0$ and thus would not contribute to the dynamics).
Eq.~\ref{eq:nonSA} can now be written in Lindblad form, thus we obtain the stated result in Eq.~\ref{eq:withSA}.

 \end{document}